# High-frequency size effect study of the Fermi surface of metals.

V.A. Gasparov, Institute of Solid State Physics,

Russian Academy of Sciences, 142432, Chernogolovka, Moscow District,

E-mail: vgasparo@issp.ac.ru.

Contents

1. Introduction
2. Ballistic effects
2.1 Cut-off of cyclotron resonance.
2.2 Radio-frequency size effect (Gantmakher effect) in a parallel magnetic field;
2.3 Studies of the Femi surface in *Cu* and *Ag*: 2.4; Time-of-flight effect in *Ag;* 2.5; Studies of the Fermi surface of transition metals *Mo* and *W*; 2.6. Radio-frequency size effect in tilted and perpendicular magnetic fields; 2.7. Multichannel radio-frequency size effect in W; 2.8 Nonlinear radio-frequency size effect in Bi.
3. **De Haas-van Alphen effect.** 3.1; Quantum oscillation in *$ZrB_{12}$*; Quantum oscillations in $YBa_2CuO_{2-x}$.
4. **Angle-resolved photoelectron spectroscopy**
5. **Conclusion**
   References

**Abstract.** **This paper reviews experimental and theoretical works on the study of the Fermi surface of pure metals with the aid of high-frequency size effects. Particular attention is given to little-known new ballistic effects such as the time of flight effect, multichannel size effect, and nonlinear size effect.**

*Devoted to the memory of Vsevolod Feliksovich Gantmakher,*

*Vitalii Vladimirovich Boiko, and Robert Huguenin.*

*Introduction*

The Fermi surfaces (FSs) of the majority of pure metals have been widely investigated since the 1960s The results of these studies are presented in a number of reviews [1-13]. To date, not only has the topology of FSs of pure metals been restored, but the basic dimensions of the FSs of these metals have also measured with high accuracy, making it possible to analytically describe the FSs of the majority of pure metals. The development of theoretical and numerical methods of construction the electronic structure of metals, such as the pseudo potential, APW (augmented plane wave) and the KKR (Korringa-Kohn- Rostocker schemes), have made it possible to conduct calculation of the electronic structure of these metals on the basis of the obtained experimental information. In spite of the fact that substantial progress has been made in *ab initio* calculations of the electronic structure of metals, the accuracy of these calculations have proven to be insufficient for the analytical description of FSs in comparison with the experimental accuracy. Therefore, various interpolation methods of calculations of the electronic structure of metals with the fitting of the necessary electron-ion potential based on experimental data were developed [14].

Subsequently, however, primary attention was given to analysis of the mechanism of electron scattering of pure metals with the aid of ballistic effects [10-13]. In this review, we have been made an attempt to gather together the existing information about the results of studies of the FSs of metals with the aid of high-frequency size effects *(HFSE),*that have not been given in the reviews published earlier [1-9].

The main purpose of this review consists in drawing the attention of physicists to very simple methods that makes possible to investigate the FSs of metals with a high precision. Notice that on the experiments on *(HFSE)* the caliper dimensions of the FSs are measured directly, which substantially simplifies the restoration of the FSs.

At the same time, in the case of effects based on quantum oscillations (de Haas-van (dHvA), effect and Shubnikov –de Haas (ShdH effect), it is the area of extremal section of the FSs that is measured. Furthermore, no very large magnetic fields or very low temperatures are required for *HFSE*, contrary to measurements of quantum oscillations. At the same time, recent experiments with quantum oscillations in high-temperature superconductors have shown the large potential of these effects for restoring the FSs of metals, so it will be given special attention in these papers.

In recent years, there have been widely developed studies of FSs with the aid of - angle resolved photoelectron spectroscopy *(ARPES).* These experiments and study procedures have been presented in a whole series of reviews [15-17]; therefore, we will consider here only the most interesting and impressive results obtained for high-temperature superconductors, referring the reader to the reviews indicated. Notice only that the accuracy of measurements with the aid of *ARPES* cannot compare with those provided by the methods of *HFSEs* and quantum oscillations.

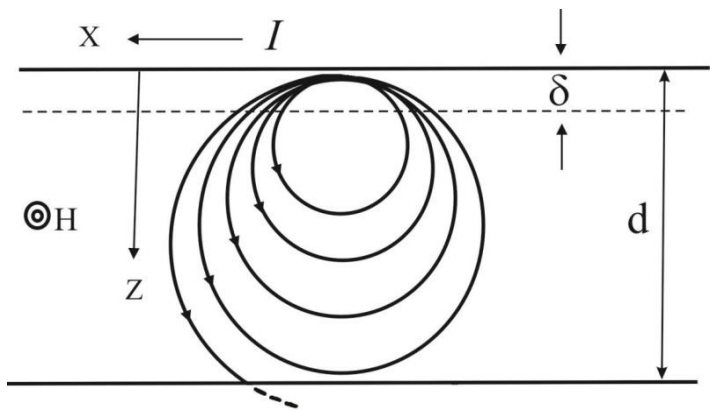

Fig.1. (Color online) Trajectories of electron motion in a sample of thickness *d* under conditions of cyclotron resonance [18].

### 1.1 Ballistic effects
### 1.2 Cut-off of cyclotron resonance.

To illustrate the HFSE phenomenon, let us examine a plane parallel single crystal-plate (Fig1) of a metal of thickness *d* with a large mean free path *l* at the liquid helium temperature in a magnetic field *H* parallel to the surface of the plate [18]. The plate is excited by microwave radiation of high-frequency (HF) field, which penetrates into the metal to the depth *δ* of the skin layer. At sufficiently high frequencies, in a sample of thickness *d* under condition of that satisfy cyclotron resonance (CR), i.e., ω=nΩ=neH/m*c, the absorption of the microwave field experiences oscillations as a function of 1/H with the Larmor frequency Ω, with the period of revolution of effective electrons that is a multiple of the period $T=2π/ω$ of the *HF* field. Here *m*\* is the effective mass of electrons (on the extremal sections of the FS), and n=1,2,3, are integer numbers. As a result of CR, resonances appear in the absorption spectrum of the microwave field in the magnetic fields $H_n=m^*ωc/ne$ with a period $ΔH^{-1}=e/m^*cω$ (Fig.2).

As can be seen from Fig.2, cyclotron resonances are observed in the thick samples that decrease in amplitude, become broadened, and pass into sinusoidal oscillations with increasing an order number *n* of the resonance. At the same time, resonances with *n* >27 disappear in the case of the thin sample in the region of magnetic fields limited by dashed straight lines. As a result, the cut-off of the cyclotron resonance occurs in the magnetic field $H_{cut}$ at which the diameter of the orbit $2ℏk=(e/c)d\ H_{cut}$ is compatible to the thickness *d* of the sample. In weaker fields, resonances can be seen from the sections with smaller diameter and mass (5 and 6 in curve 2). At the same time, it should be noted that no special features in the field of the cut-off field $H_{cut}$ were observed. From studies the position of cut-off field $H_{cut}$ and of the period of oscillations of the cyclotron resonance depending on the orientation of the magnetic field, the FS of *Tin* in the first Brillouin zone (BZ) in the plane (100) and the anisotropy of cyclotron mass of electrons *m*\**(k)* in this section have been restored [18,19]. Subsequently, however, cut-off of the CR was observed only in *In* [20], in spite of the detailed CR studies performed in **Bismuth** [8,9], **Al** [21], and **Lead** [22].

*1.1 Radio-frequency size effect (Gantmakher effect) in a parallel magnetic field.*

A true breakthrough in studies of the FSs in metals with the aid of *HFSE* was made by V.F. Gantmakher [23] in experiments using radio frequencies in **Tin**. The studies were carried out under the same conditions as in the work [18,19] on CR, but at frequencies that were four orders of magnitude lower (1-5) MHz. No CR was observed at these, so low frequencies. From the equation of motion of electrons in the magnetic field with respect the time *t*

$$ℏ\dot{k} = (e/c)[v × H] \quad (1)$$

it can easily be shown [5,18,19, 23-27], that the orbits of electrons in the momentum and real spaces are turned relative to each other through the angle π/2. In this case the relationship between the diameter *2ℏk* of the orbit of electrons in the momentum space and the thickness *d* of the plate makes it possible to determine the diameter $2\hbar k_F$ of the section of the FS perpendicular to $H$. This diameter of the section of the FS in the momentum space is equal to:

$$k_F = \left(\frac{e}{2\hbar c}\right)H_0 d \qquad (2)$$

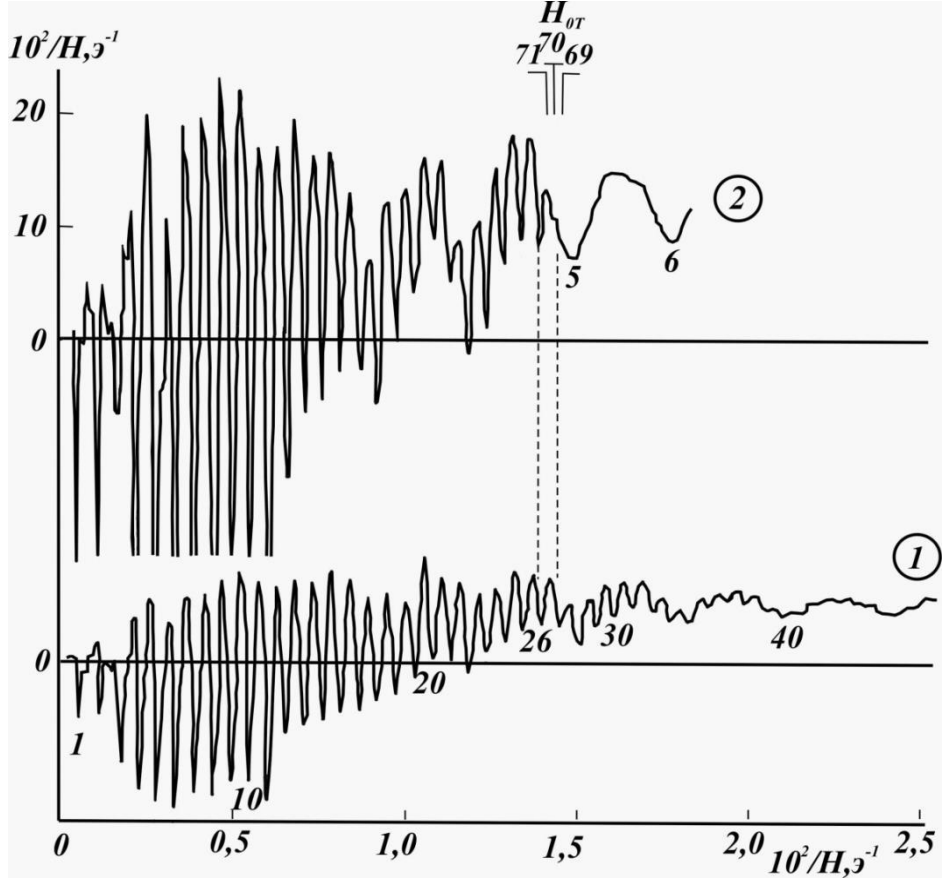

Figure 2. (Color online) Derivative ∂f/∂H of the frequency of a strip resonator with respect to the magnetic field recorded on single crystal-Sn plates with a thickness of 2 mm (curve 1) and 0.482mm (curve2). It can be seen that in the thinner sample the higher-order (27 and above) resonances are cut off (Sn, T=3.75 K, ω=9.4 GHz, H//C4, [18]. $H_{cut}$ is a cut-off field of the cyclotron resonance: [69, 70, 71]

Studies of the position of *RFSE* lines in the field $H_0$ on the orientation of the magnetic field made it possible to restore the section of the FS of *Tin* in the first Brillouin zone (BZ) [23-26] (Fig. 3b) in excellent agreement with data on the cut off of the CR [18,19].

In further experiments, it was discovered, that besides the basic line of RFSE in the field $H_0$, additional lines are observed in the fields that are multiples of $H_0$ i.e., $2H_0$, $3H_0$, etc. In the multiple magnetic fields, the electrons excite the skin layer deeply in the metal at a distance that is a multiple of the diameter $2\hbar k_F$ of the orbits, when the component $v_z$ of the velocity of elections changes its sign upon reaching the bottom of the orbit [26]. As a result, splashes of the HF field with a phase that is opposite to the phase of the initial skin layer are excited in the bulk of the metal. In experiments on *Tin* [23], the electrons move along the cylindrical section of the FS, therefore, the contribution from different sections reaches its maximum. In the doubled field $2H_0$, the electrons excite splashes of *HF* field at a depth of *d/2*, which serves as a source of a new splash in the field $3H_0$, etc. see Fig.4. As a result, splashes of the HF field appear at the depth of the metal $2r_{extr} = 2\hbar k_{extr}/eH$, while it's well screened from HF excitation in the zero fields. The structure of the splashes of the field deep within the metal, $E=E(z) \exp - (i\omega t)$, is described by the equation [28].

$$\frac{E(n)}{E(0)} = \frac{\Gamma\left(\frac{3}{4}\right)}{\pi\sqrt{2}}(-1)^n \cos\frac{n\pi}{4} a^n \frac{\Gamma\left(n+\frac{1}{4}\right)}{\Gamma(n+1)}$$

$$\Gamma(z) = \int_0^1 (-\ln x)^{z-1} dx \qquad (3)$$

Here $n$, is the number of field's splashes in a plate of thickness $d$, $\delta o$, $\delta_1$, and $\delta_2$ are the depth of the initial skin layer, of the nth splashes and the splash near the boundary $z=d$, respectively (see Fig 4 [27], $a$ is the amplitude of the **n**th splash and $\Gamma(z)$ is the gamma function.

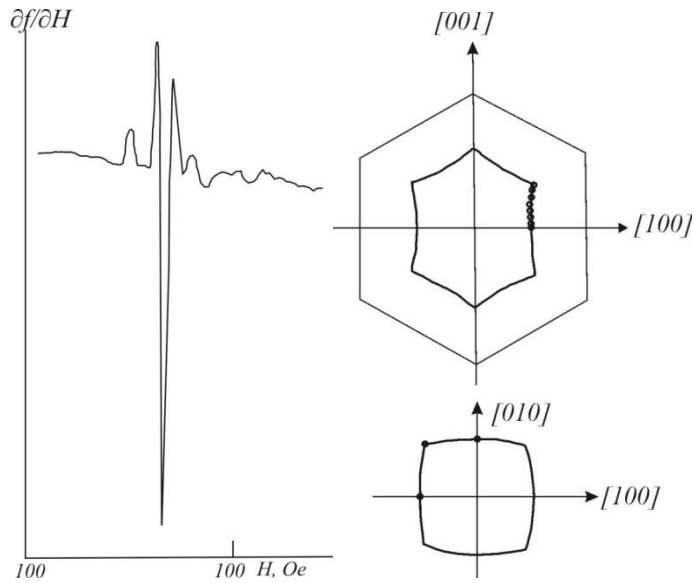

Figure 3.(a) (Color online) Derivative (with respect to the magnetic field) of the real part of the impedance of a **Tin** sample *(d=0.54mm)*, recorded at n//[001], H//[001], T=3.75K, and f=2.8 MHz (b) Cross section of the FS of **Tin** in the fourth zone in the model of almost free electrons. Points correspond to the results of studies of the *RFSE* in a parallel magnetic field [23-25].

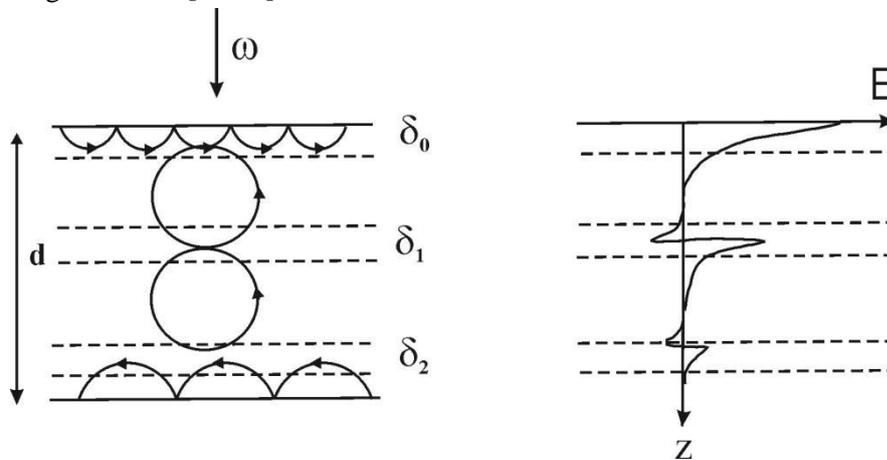

Figure 4. (a) (Color online) Electron orbits, and (b) the distribution of the HF field in the bulk of the metal under *RFSE* in multiple magnetic fields [28].

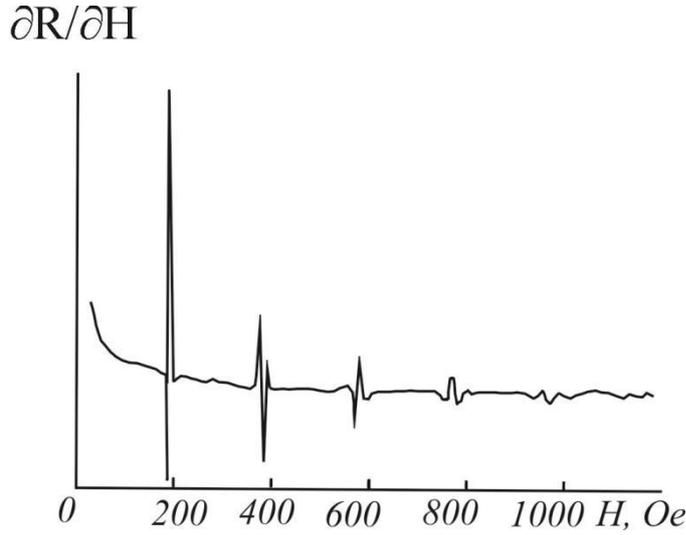

Figure 5. (Color online) RFSE lines recorded in multiple magnetic fields in *Ag* (d=0.805mm, T=4.2K, ω/2π=2.2 MHz, n//***<110>, H//<100>***) [29].

The amplitude of the basic RFSE line in the field *Ho* is determined by the probability of electron scattering during orbital movements. Under the condition of the extremal anomalous skin effect (l<<d), the contribution to the amplitude comes from an exponentially small number of electrons. This makes it possible to study different mechanisms of electrons scattering rate $\bar{\nu}$(T) (electron-phonon, electron-electron, etc.) based on the temperature dependence of the amplitude of RFSE line [13]:

$$A \propto \exp\left(-\frac{\pi\bar{\nu}}{\Omega}\right) \quad (4)$$

Here $\bar{\nu}$ is the electro-phonon, electron-electron scattering rate of electron averaged along the orbit, and Ω is the cyclotron frequency:

$$\bar{\nu} = \frac{\hbar}{2\pi m^*} \oint \frac{\nu(k)dk}{v_\perp(k)}, \quad (5)$$

A review of these experiments is published in [13] and is beyond the scope of this paper where we will be mainly interested only in studies of the FSs with the aid of *HFSEs*.

The value of the field $H_0$ in Eq. (2) is determined by the position of the *RFSE* line, depending on the magnetic field orientation, which is an essential question in calculations of the radial size $k_F$ of the FS. This problem was solved in papers [29,39], from the study of the dependence of the position of different extremes of the RFSE lines in *In* and *Mo* on the frequency *ω* and thickness *d* of the sample, respectively. As follows from Fig. 7, a change in the position of different extrema of the *RFSE* line depends on the thickness *d* of the sample in *Mo* asymmetrically. This feature makes it possible to determine the field $H_0$ in equation (2) from the left-hand edge of the RFSE lines [29, 30], which in turn allows measuring the cross-section diameter $2\hbar k_F$ of the **FS** of *Ag* with high precision (see Section 2.5). From similar measurements using rather thick samples of **Cu** (*d= 3mm*) we have shown that the sizes of the FS determined from left-hand edge agree with the data on the *dHvA* effect with an accuracy of ±0.2%. [29-32].

### 2.3. Studies of the Fermi surface in Cu and Ag.

As an example of the potential of the RFSE in a parallel magnetic field for restorations of the FSs, let us examine some results of investigations of this effect in **Cu** and noble metals (**Ag**), where the potential of the *RFSE* for determination the sizes of the FSs manifest itself with high precision.

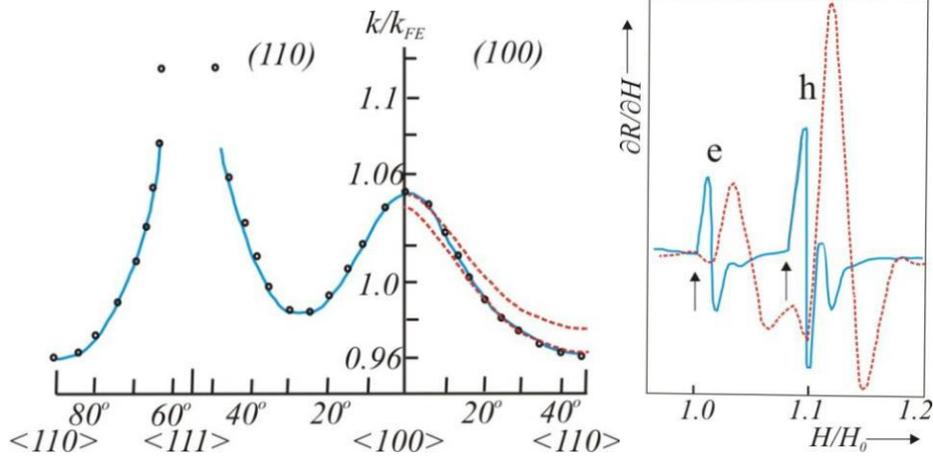

Fig. 6. (Color online) Anisotropy of the positions of *RFSE* lines in a parallel magnetic field in **Ag** in the planes (110) and (100) [29]. The red dashed lines shows the result of *ab initio* calculations of the FSs at different values of $E_F$ [35]. The circles correspond to the *dHvA* data [34]. The solid curve – *RFSE* results approximated according to Eq.6 [29].

Fig. 7. (Color online) RFSE lines in **Mo** using samples of various thicknesses *d=0.955 mm* (solid curve), and *d=0.245 mm* (dashed curve). The x- axis is normalized with respect to $H_0$ for the line from electron jack at *Γ (e)*. The arrows show the positions of the lines for the jack (e) and hole octahedron at *H*(h) (see Section 2.5 [29].

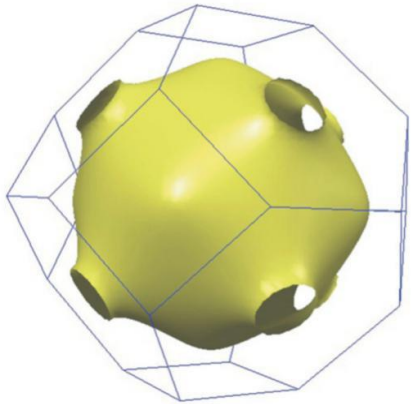

Figure 8. (Color online) FS of noble metals, (http://www.phys.ufl.edu/fermisurface/)

The Fermi surface of the noble metals is presented schematically in Fig.8 [34]. This surface is open along the axis <111> and occupies half of the BZ, whereas in the model of free electrons it takes the form of the sphere inscribed in the BZ. The anisotropy of the position of the *RFSE* lines in a parallel magnetic field in the planes (100) and (110) in **Ag** is illustrated in Fig.6. [34]. On the ordinate axis, the radius vector of the FS is plotted in units of the radius vector of the sphere of the free electrons $k_{FE}=1.2068$Å$^{-1}$ at 4.2K:

$$k(\theta, \varphi) = (99.58 - 0.16\, K_1 + 1.63\, K_2 + 1.36\, K_3 - 0.53 K_4 + 0.03 K_5 + 0.29\, K_6) \times 10^{-3} k_{FE} \quad (6)$$

The solid curve in Fig.6 shows the fitting of $k(\theta,\varphi)$ experimental data in the form of a series of cubic harmonics [29] according to equation (6). Here **k**, **θ** and **φ** are the spherical coordinates of a point on the FS with a high accuracy (±2%), except for the region of the neck, which in the experiments on the RFSE in Ag was not observed because of very high electron-phonon scattering rate in this region of the FS [29,33]. As can be seen from Fig. 6, the measured sizes of the FS in **Ag** are in excellent agreement with the data of the dHvA effect [34], which indicates the high accuracy of *RFSE* method. A similar agreement of the data of the *RFSE* and dHvA effect was demonstrated for **Cu** [31].

There is a large amount of works in which an attempt was made to describe the FS of noble metals from the first principles calculations [14]. Nevertheless, as can be seen from Fig. 6 in spite of the wide development of *ab initio* methods, the theoretical models, although describing the shape and sizes of FSs, do not lead to a better agreement with the experiment. In this sense, a promising method is the utilization of the parametrization of the lattice potential with the fitting of the parameters according to the available *RFSE* experimental data for FSs [14]. Apart from the above-described examples of employment of the *RFSE* for studying FSs, numerous studies have been made of the anisotropy of the electron-phonon and electron-electron scattering rates, depending on the position on the FS in **Cu** and **Ag** [13,29,33,36-38].

### 2.4. Time-of-flight effect in Ag.

Comparatively, little attention has been given to another *HFSE* – the time of flight effect (*TFE*) [38]. This effect was firstly discovered in **Ag** in experiments in a parallel magnetic field in the configuration adopted in the *RFSE* studies, but at four-orders-higher frequencies (45 GHz). In these experiments, one side excitation of the **Ag** plane-parallel plate was used under conditions of cyclotron resonance. As can be seen from Fig. 9, no *CR* was observed in magnetic fields higher than the cutoff field $H_0$, since the diameter of the electron orbit exceeds the thickness **d** of the sample. However, a microwave signal passing through the sample experiences oscillations, depending on the magnetic field parallel to the sample surface.

A theory of this effect was developed in the paper [38]. The main idea of the *TFE* consists of the interference of the weak signal of the leakage around the sample with one along a ballistic trajectory. As a result of the oscillations of the time of flight relative to the phase of the signal of leakage, an oscillatory dependence of the signal having passed through the sample on the magnetic field appears (see Fig.9). As follows from Fig.9, the position of the *RFSE* line in the field $H_0$ determined according to the left-hand edge side agrees well with the cutoff field $H_0$ of the *TFE*, in accordance with the frequency dependence of the *RFSE* extrema in **In** [30] and with the dependence of the thickness of the sample in **Mo** [31]. As was shown by V.F. Gantmakher *et al* [39], the main contribution to the *TFE* comes from type **2** orbits (Fig. 10), for which the time of motion in the skin layer ($\delta/r$) is a maximum, as opposed to type **1** and **3** orbits, and correspondingly, the influence of the effect of the time delay is minimal. The time of flight in orbit **2** can be represented in the following form [40]:

$$t = \frac{2\hbar}{eH} \times \int_0^\emptyset \frac{k d\emptyset}{v_\perp \cos\chi} \quad (7)$$

The amplitude of the TFE oscillations is proportional to the number of electrons having passed without scattering along orbit **2**, and takes on the same form as amplitude of the *RFSE* line in the parallel magnetic field (formula *5*), with the difference being that $\bar{v}$ has been averaged over a segment of orbit **2**, rather than over the entire orbit, as in the case of the *RFSE* [40, 41]:

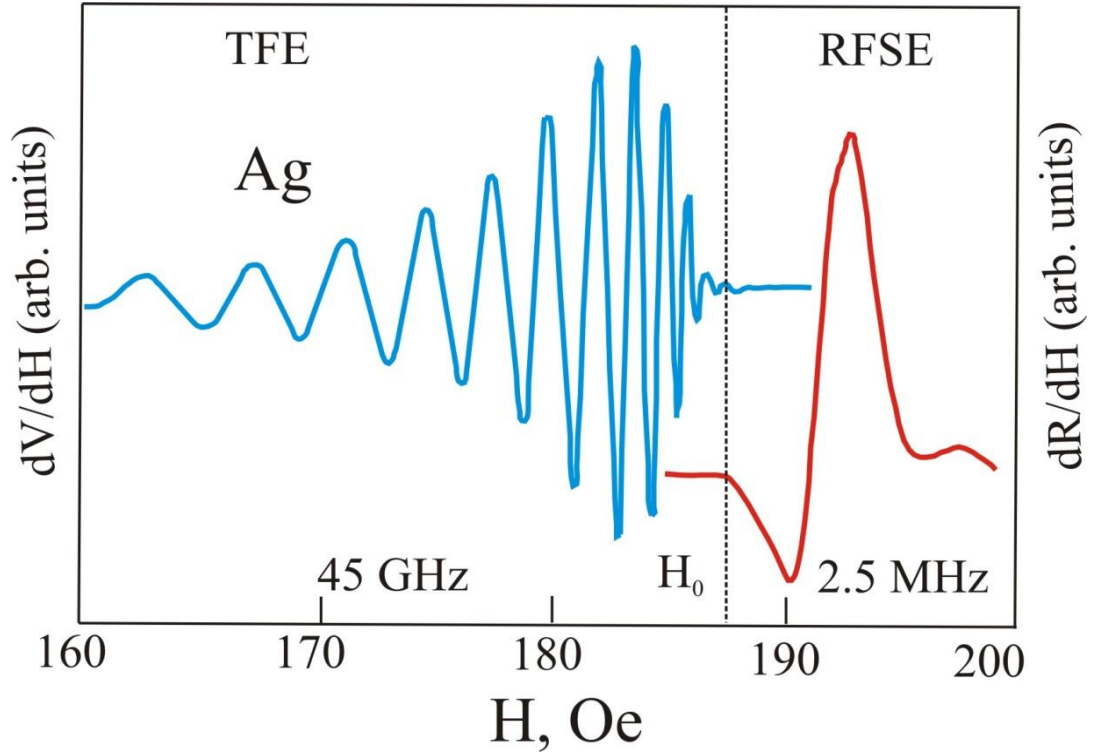

Figure 9. (Color online) RFSE in silver recorded at 2.5 MHz, and TFE at 45 GHz. (n//[110], H//[001], d=0.811 mm) [40,41].

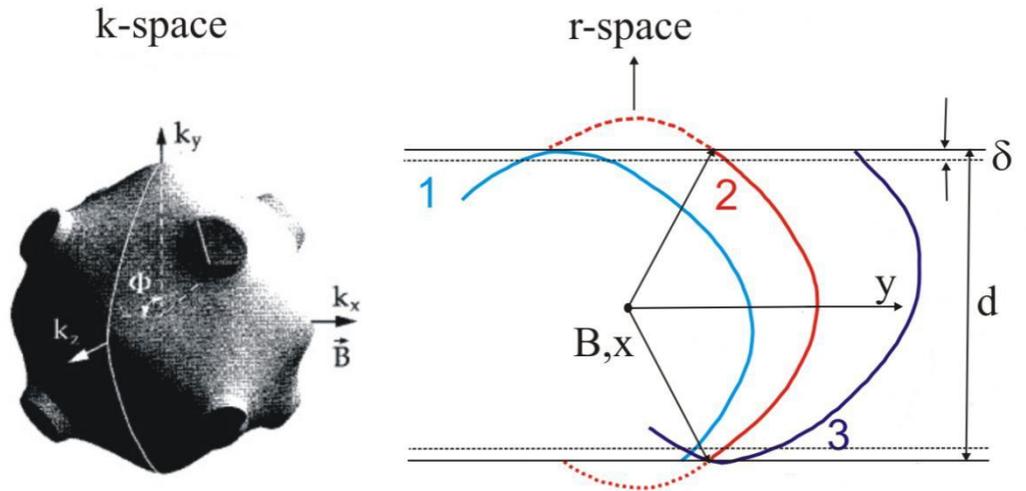

Figure 10. (Color online) Orbits of electrons in the *k* and *r* spaces that make a contribution to the TFE in Ag. The magnetic field is parallel to the surface of the plate and to the <100> axis.

$$\bar{v} = \int_0^\phi \frac{v(k)k d\phi}{v_\perp(k)\cos\chi} \Big/ \int_0^\phi \frac{k d\phi}{v_\perp(k)\cos\chi} \qquad (8)$$

Here, $X$ is the angle between $v_\perp$ and **k**, and $\varphi$ is the angle of the start of a flight from the surface. Expression (8) contains information about the velocity of electrons averaged over orbit *2*, which makes it possible to restore the Fermi velocity $v(\mathbf{k})$ on the FS from the position of oscillations of the signal passed through the sample, versus $H$ [39,40]. It can be shown [39,40] that the relation between the Fermi velocity $v(\mathbf{k})$ of electrons and the experimental dependences of the position of *TFE* oscillation is determined by the following equation:

$$v_x = \frac{\omega D}{H\left(\frac{d\phi}{dH}\right)+\phi} \qquad (9)$$

In fact, only a change in the phase $\Delta\varphi(H)$ is measured in experiments, where the phase $\varphi = \pi H_c/H_0$ is determined by the cyclotron mass known from the CR ($H_c=\omega m^*c/e$), $H_0$ is the field of *TFE* cut-off, and *D* is the diameter of the orbit in the momentum space. As a result we obtain

$$\phi = \Delta\phi(H) - \Delta\phi(H_e) + \pi H_c/H_e \qquad (10)$$

Equation (8) makes it possible to determine the local frequency of electron-phonon scattering rate [41,42] with the aid of relationship (7) for the time of flight of electrons [39,40]:

$$t(H) = \frac{2\hbar}{eH}\int_0^{k_y}\frac{dk}{v_x}, \quad k_y = \frac{eHD}{2\hbar c} \qquad (11)$$

In reality, however, the accuracy of this calculations procedure for the inversion frequency of e-p scattering *v(k)* and for the velocity υ(k) is somewhat limited because the presence of different corrections [40], and also because of the insufficient accuracy of the FS model, necessary for calculations. At the same time, the *TFE* proved to be a sufficiently efficient method for restoring the velocity of electrons [39,40] and the frequency of electron-phonon scattering rate *v(k)* on the FS of *Ag* [41,42].

**2.5 Studies of the Fermi surface of transition metals *Mo* and W.**

Another example of the successful application of the *RFSE* for restoring the FSs of metals are studies of the metals of the chromium group (*Mo, W*) [31,32,45-51]. Figure 11 shows the Lomer model of the FS for these metals [42,43]. The electron sheet consists of the electron *Jack* at the point *Γ* of the Brillouin zone, at the corners of which the six electron balls are located.

The hole sheet, consists of an octahedron at the point *H* and six ellipsoids at the points *N* (green points in Fig.12). Inside the necks between the *jack* and the hall parts of the FS in *Mo* are small electron pockets, which are absent in *W*.

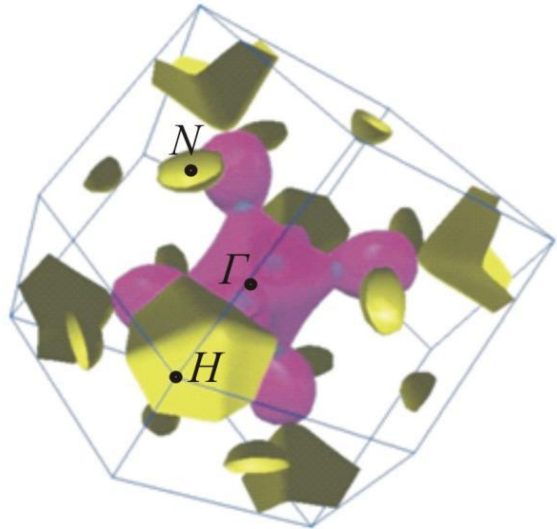

Figure 11. Fermi surface of chromium-group http.phys.ufl.edu/fermiscurface/.

Figure 12a displays polar diagrams of the extremal sizes of the FS in *Mo* restored from the anisotropy of position of *RFSE* lines in the plane (011) [31,32,45-52]. The wave vector **k** is plotted versus the radius vector in the units of $k_0 = 2\pi/a = 1.998\text{Å}^{-1}$ (the size of the BZ along the [100] axis). Here, $a=3.144$ Å is the lattice parameter of *Mo* at liquid-helium temperature.

The position of the *RFSE* lines in the field $H_0$ was determined from the left-hand side of the *RFSE* lines [31.

The error in measurements didn't exceed 1%. At the same time, the presence of a large number of closely spaced lines, which intersected with the lines in multiple magnetic fields, strongly impeded their identification for different sheets of the FSs. Nevertheless, all the RFSE lines turned out to be identified [44,48,45-52]:

1. Hole octahedron at the point **H**. As can be seen from Fig. 12a, the **h** lines well describe the shape of the section of the hole octahedron of the FS in **Mo** with the symmetry center at the point **H** of the BZ (see Fig.11). This circumstance made it possible to determine all the main sizes of the hole octahedron in **Mo** and **W**: the area of the surface in **Mo** S=5.0 Å$^{-2}$, and the volume V=0.61 Å$^{-3}$[46-50].
2. Hole ellipsoids at the point **N**. Sections $d_1, d_2,$ **and** $d_3$ in Fig. 12b well describe the sections of the hole ellipsoids at the point **N** of the BZ. These ellipsoids are elongated along the [100] axis and are compressed along [110]. Investigations in different planes – (100), (110), (111), and (121) – made it possible to determine the dimensions of the principal semi axes **a, b, and** **c** of these ellipsoids with a high accuracy owing to the use of relatively thick **Mo** and **W** samples [31,47,48]. The total area of the surfaces of ellipsoids in **Mo** is S=6.6 Å$^{-2}$, their volume is V=0.61 Å$^{-3}$.
3. Electron **jack** at the point $\Gamma$ and the electron **lenses**. Sections **i** and **g** primarily refer to the FS sections on this sheet (Fig.12a).
4. Apart from these lines, the following lines were also observed: **e**, *spheroids*, **b,** *necks* and the octahedron part of the **jack**.
5. Section **y** is due to the kink in the orbit **i.** This made it possible to determine the distance from $\Gamma$ to the plane of the neck [46]. The sections $a_1$ and $a_2$ of the neck well describe the cone elongated along the [100] axis with a rounded base and an apex. The anisotropy of the position of the lines **e1** and **e2** makes it possible to relate them to electron spheroids. The radius of the spheroids is determined by the section $e_1$ along the [001] axis with a radius vector equal to 0.032 Å$^{-1}$. Lines **c** and **b** are due to kinks in the orbits of electron **jack** [31,47,48].

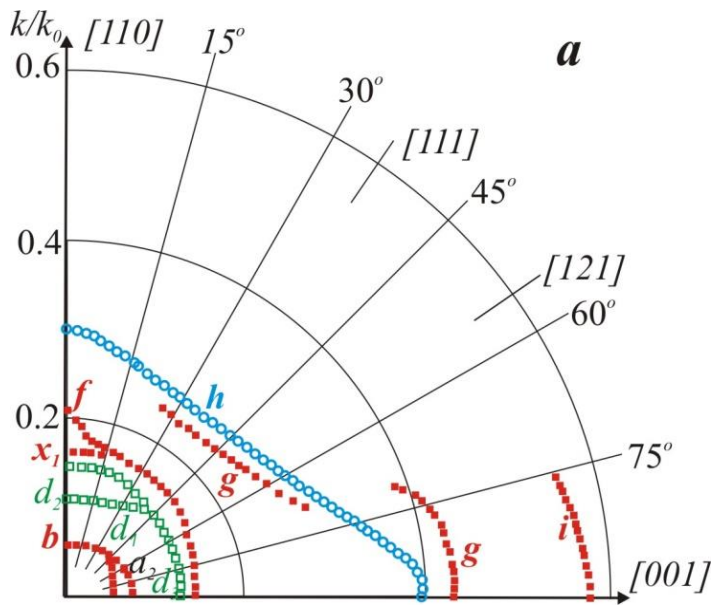

Figure **12a**. Color online) (a) Angular dependence of the wave vector k/k$_0$ (k$_0$=2π/a, *a*=1.998 Å) in the plane (110) in **Mo** [31,32,42,45-53]. The red squares correspond to the sections from electron surface – jack (g); blue circles – to hole sheets (h), green squares to hole ellipsoids (*d1-d3*).

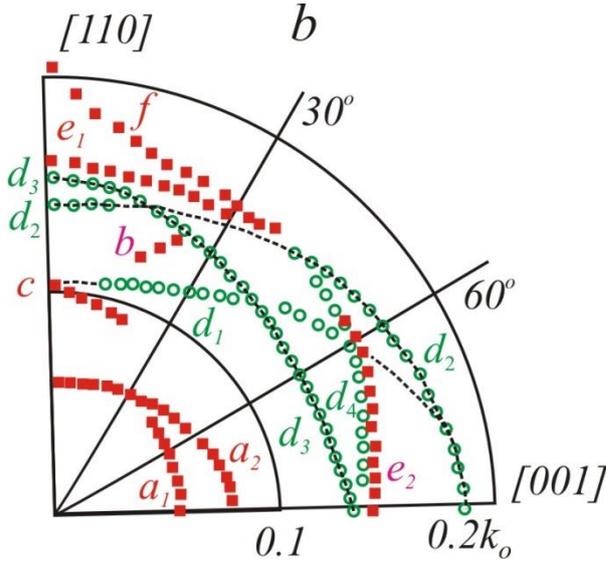

Figure 12**b**. Color online) Angular dependence of the wave vector $k/k_0$ ($k_0=2\pi/a$, $a=1.998$ Å) in the plane (110) in ***Mo*** [31,32,42,45-53]. Notations are the same as in figure (a) but at small $k_F$.

Figure 13. (Color online) Empirical model of the FS in ***W*** (solid lines) and ***Mo*** (dotted lines [48]). An analyses of the positions of RFSE lines made it possible to describe the all sections of the FSs in ***W*** and ***Mo*** [46-48].

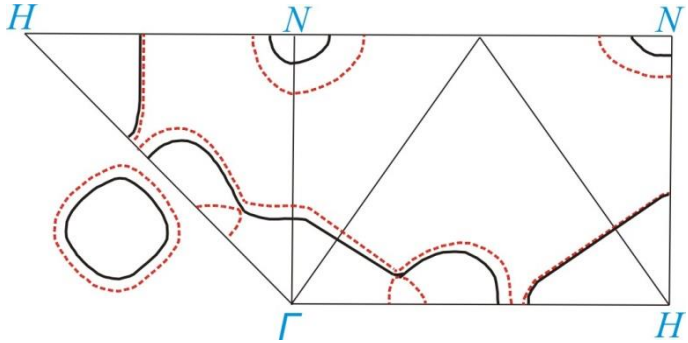

Angular range of the existence of lines ***b*** and their anisotropy makes it possible to relate them to the orbits on the neck of the ***jack*** connecting the octahedral part of the ***jack*** body with the spheroids. An empirical model of the FSs in ***Mo*** and ***W*** [48] restored from experiments on the *RFSE* in the parallel magnetic field is shown in Fig.13. An analysis of the positions of these lines made it possible to describe the sections of the FSs in ***W*** and ***Mo*** with a high accuracy [46,47].

**2.6 Radio-frequency size effect in tilted and perpendicular magnetic fields.**

As was already noted above, the *RFSE* in parallel magnetic field is due to "effective electrons", which move in the skin layer parallel to the surface with a velocity $\upsilon_z=0$. Another example of *RFSE* on "effective electrons" is the *RFSE* at the limiting point or at the section of the FS with an extremal derivative of the surface area of the section with respect to the momentum, $\partial S/\partial k_z$ [5,13,30,51-54,56]. In this case, the electrons start moving in the skin layer parallel to the surface and continue moving deep in the metal in spiral orbits (Fig. 14) [13,27,30]. The extremal shift of electrons deep to the metal (***u***) depending on the electron momentum $k_z$ along the magnetic field can take place for the orbit at the elliptic limiting point [30, 54, 56]:

$$u_0 = 2\pi c/eHK^{1/2} \tag{12}$$

Here $K$ is the Gaussian curvature of the FS at the limiting point of the FS.

For an arbitrary FS, such extrema can occur at the no central sections of the FS with an extremum of $\partial S/\partial k_z$ (see Fig. 14) [5, 13, 30, 54-58]:

$$u_1 = \frac{c}{eH\hbar}\left|\frac{\partial S}{\partial k_z}\right| \tag{13}$$

where, $S(k_z)$ is the area of the cross section of the FS with the plane $k_z$=const.

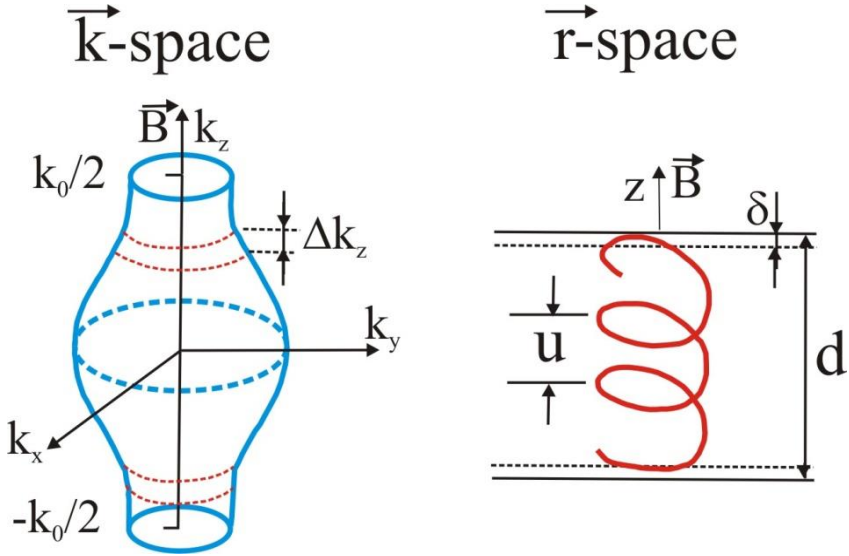

Figure 14. (Color online) Orbits of electrons in the **k** and **r** spaces for the corrugated FS, which are responsible for GKO effect in the magnetic field perpendicular to the sample surface [13].

The electron motion in such an orbit in a parallel magnetic field leads to a shift of the position of the *RFSE* lines according to the law:

$$H = H_0 \times cos\varphi \tag{14}$$

for the *RFSE* from the orbits at the noncentral section of the FS [30,54,55]. Here $D$ is *a* diameter and $h$ is the tilt angle of the field, and $H_0$ is the position of the *RFSE* line in the parallel magnetic field. For electrons on the no central sections of the FS in a tilted field, a drift component $\upsilon_H\varphi$ of the velocity oriented along the normal to the surface **n** appears. It can be shown that the diameter $D$ of the spiral orbit is related to the plate thickness *d*:

$$d = D \cos\varphi = \left(n - \frac{1}{2}\right) h \sin\varphi \qquad n = 0,1,2,\ldots \tag{15}$$

Here, $D$ is a diameter, and $h$ *is* spiral orbit pitch on the non-centrally section of the FS.

$$\Delta H = H_n - H_{n-1} = \frac{2\pi c}{ed} K^{-1/2}\, \varphi \tag{16}$$

The period *ΔH* in the position of these lines made it possible to determine the Gaussian curvature $K$ of the **FS** in ***In*** at the limiting point of the FS [30,54].

The RFSE in the case of limiting point orbits or on the non-centrally section of the FS take place at those values $H_n$ of the magnetic field for which the number *n* of rotations along the path from one side of the sample to the other side is an integer. As a result, *RFSE* lines were observed at the limiting point in ***In*** for the direction of the magnetic field along [111], which were periodic with respect to the field [30, 54-56].

The RFSE in the case of the non-centrally sections on octahedron sheet of the FS in *Mo* and *W* were observed as well [48,48] (Figs 15a, and 15b). The same effect is responsible for the splitting of RFSE lines in *W*, in the tilted magnetic field [48]. However, the variety of RFSE is not only due to the narrow character of the lines in the parallel or tilted magnetic field. As follows from Fig.15b, harmonic oscillations of the derivative of the surface impedance in *Mo*, in a magnetic field perpendicular to the surface, were observed with respect to the magnetic field [48]. Similar oscillations were observed in a perpendicular field in *Sn* [55] and *Cd* [55-56]. This effect is known as the Gantmakher-Kaner effect [54], and is due to the existence of ineffective electrons with a nonzero component $v_z$ of the velocity along the normal *n* (see Fig.14).

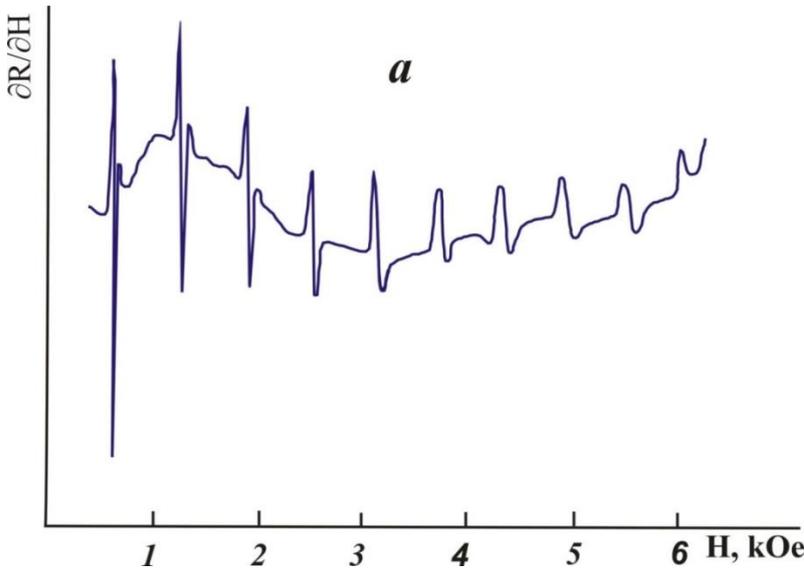

Figure **15a.** (Color online) RFSE lines record on a tilted magnetic field in *Mo* and caused by the orbits on a hole octahedron in the point *H BZ* with an extreme derivative $\partial S/\partial k_z$ [48]. Tilt of the magnetic field is in the plane ([110], n, d=0.245mm, f=3.5MHz, and T=4.2 K).

As follows from Fig.15b, harmonic oscillations of the derivative of the surface impedance in *Mo*, in a magnetic field perpendicular to the surface, were observed with respect to the magnetic field [48]. Similar oscillations were observed in a perpendicular field in *Tin* [55] and *Cd* [55,56].

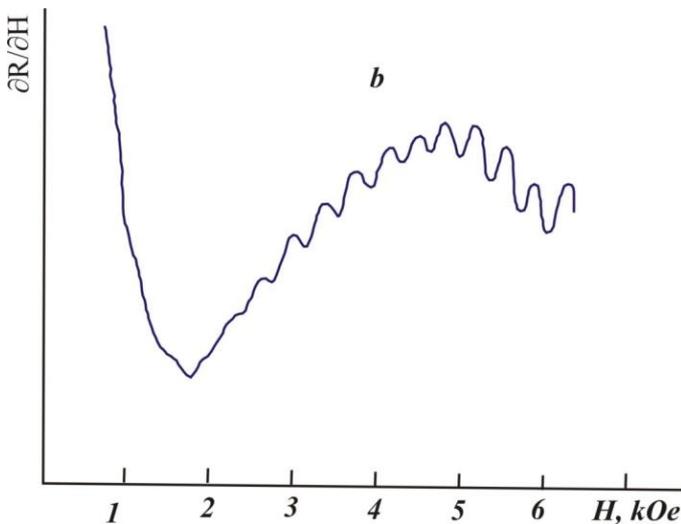

Figure **15b** (Color online) RFSE lines recorded from the no central sections on the hole octahedron in *Mo* in a perpendicular magnetic field (n//[110], f=3.5 MHz, T=4.2 K, d=0.239 mm) [47].

The excitation of electrons by a radio-frequency field are the same, irrespective of the orbit, for those harmonics of the RF field in metal for which the wavelength $\lambda_n = u/n$ (n=1,2,3…) is a multiple of the shift $u$ of the electrons into the metal per cyclotron period [4,54,56].

This feature leads to a harmonic distribution of the RF field in a metal, contrary to the case of geometrical resonances considered above for the cases of parallel and tilted magnetic fields. In the harmonic case, the metal becomes transparent to the electromagnetic excitation. The period of Gantmakher-Kaner oscillations (GKOs) is determined by the extremum of the displacement $u$, just as in the case of *RFSE* in a tilted magnetic field [13] which made it possible to determine such an important characteristic of the FS as the angular dependence of the derivative $|\partial S/\partial k_z|$ from experiments in both the tilted and perpendicular magnetic field in *Mo* [47].

### 2.7. Multichannel radio-frequency size effect in W.

Up to now, we have considered the *RFSE* in the case of a "normal" scattering of electrons from the surface of the metal, when the electrons remain on the same sheet of the FS of the metal after surface scattering. However, the law of momentum conversation upon surface scattering requires the conversation only of the tangent component $k_{\parallel} k_n$ of the wave vector of electrons, where *n* is normal to the surface of the sample. Therefore, as follows from Fig.16, surface scattering in the case of multi sheet FS can occur with a transfer between different sheets of the FS or even between the different BZ (umklapp scattering. Such character of the surface scattering was considered theoretically in a whole number of papers [59-63].

If the plane of the orbit lies near the limiting point of the sheet of the FS, then the electron moving along the jumping trajectories from A to B on the electron jack (see Fig.16) can experience not only conventional specular scattering to point B but also perform jumps to point C (see Fig. 16) on the same sheet of the FS with the same $k_{\parallel}$.

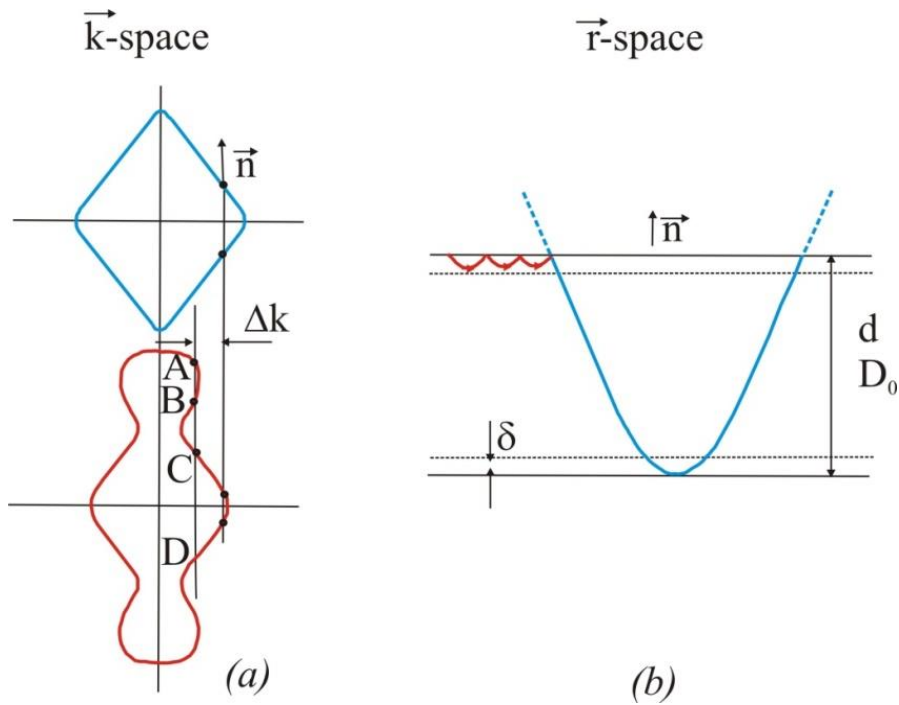

Figure 16. (Color online) Multichannel surface scattering of the electrons in the *k* and *r* spaces in *W* and *Mo*.

Such electrons can continue moving in the jumped orbit and even jump from electron sheet onto a neighbour hole sheet of the FS (see Fig.16). As a result, a hole orbits can appear with a larger diameter of the orbits and, correspondingly a new *RFSE* line will appear. The probability of these jumps is rather low because is proportional to the angle of scattering $\varphi$ [63-65]:

$$p_{BC} = \alpha_{BC} v_B(k_{||}) v_C(k_{||}) = \alpha_{BC}(k)\varphi \tag{17}$$

Here, $a_{BC}$ depends on the wave functions of electrons at the points **B** and **C** and $v_B$ and $v_C$, are the normal components of the electron velocities at these points.

As can be seen from Fig.17, besides of the RFSE lines invesigated earlier [47], new lines were observed in the magnetic field of 30 Oe in **W**, which disapears after etching the surface of the sample. At least two new *RFSE* lines were also observed in a samples with mirror-finished surface [64] in the experiments at H//[100], which were caused by different types of jumps of holes jumping onto electron sheroid and back, and also of electrons jumping onto the spheroid in the adjacent BZ [64]. The position of these lines agrees well with the jumps upon multichannel scattering. A change in the frequency did not affect the position of these lines. They disappred upon a change in the orientation of the magnrtic field by several degrees relative to the optimum orientations.

The decrease in the sample thickness lead to the displacementn of these lines into the range of larger fields together with the basic RFSE lines (dashed cuves in Fig.17) [64]. The amplitude if these lines is approximately an one order of magnitude lower than the amplitude of conventional lines in closed orbits because of small probability of specular scattering for the jumping orbits, which is proportional to the incidence angle (see Eq. 17.

As can be seen from Fig. 16, there are a large number of possible channels in **W** and **Mo** for multichannel surface scattering. We observed this type of RFSE in a parallel magnetic field in **W** samples with mirror-finished surface [64]. Any of these jumps can lead to the appearance of new *RFSE* lines in different magnetic fields, which were not observed on the samples with difuse surface. Apparently, it is because of the small probability of the surface specular scattering that the latter was not obseved in experiments on electron focusing in the tranferse magnetic field in **W** [65]. This method is very efficient for study the specular reflection of electrons in metals. [66].

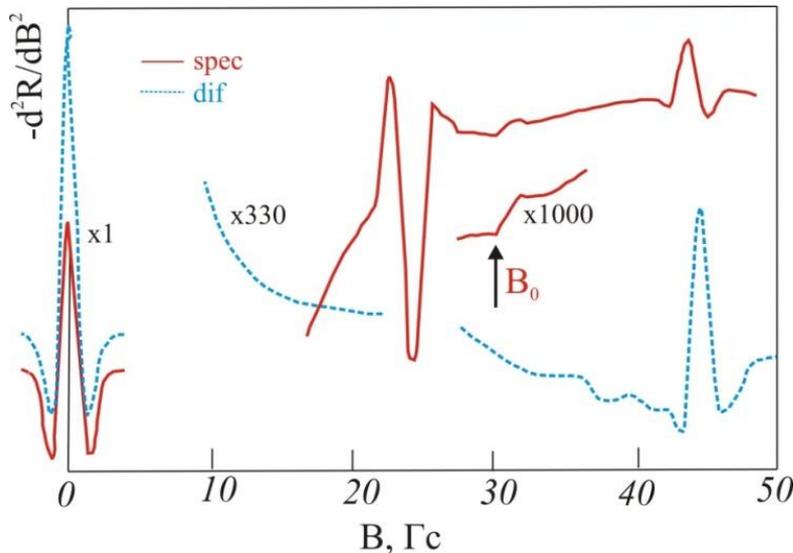

Figure 17. (Color online) RFSE lines recorded in a W sample with a specular (solid lines) and diffuse (dotted lines) surface, H//[100], d=0.43mm,$\omega/2\pi$ = 1.9 MHz, T =1.35K.

**2.8 Nonlinear radio-frequency size effect in Bi**

So far we have considered the *RFSE* effect under linear conditions, when the amplitude of RF field $H_\omega$ of the radio-frequency field on the surface of a sample is small compared to an external DC magnetic field $H_0$. In this case, the electron orbits are not distorted in the skin layer. The situation changes substantially in the case of the reverse limit, when $H_\omega \geq H_0$ [69,70]. Under these conditions, the magnetic field of a wave superimposed on $H_0$, leads to the non-equivalence of the two half-periods of the alternating current in the skin layer. As a result, a DC component of the RF current arises near the surface. This current attenuates at the depth $\delta$ but is carried into the depths of the sample in the form of ant symmetric splashes of the direct current with an alternative sign, just like splashes of the HF current in the *RFSE* through the chain of orbits (see Section 2.2, Fig.4).

The amplitude of the splashes diminishes with the order number *n* according to the law $n^{-2/3}$, almost just as the splashes of the HF field dumping [(Eq. 3)]. The magnetic field of the rectified current acts on the orbits of electrons together with an external magnetic field. As the external field reaches the strength $H_n = 2nvmc/ed$, where *d* is the thickness of the plate, and *n* is an integer, the splashes of the direct current go into the opposite surface of the sample and can be detected from the dependence of magnetic moment of the sample on the external magnetic field [69-73]. The *RFSE* under nonlinear conditions was observed in **Bi** [70,72] from the dependence of the derivative $dM/dH$ of the magnetic moment of the sample versus DC magnetic field parallel to the surface of the sample (Fig.18). Actually, the FS of the semi metallic **Bi** consists of three electron sheets and one hole sheet close in shape to ellipsoids with anomalously small sizes in comparison with the metals examined above (see inset to Fig.18). Therefore, the linear *RFSE* is observed in a very weak magnetic field of 1-2 Oe. At larger amplitudes of the HF field, from 0.5 to 6 Oe, a, nonlinear *RFSE* was observed from the dependence of the derivative $dM/dH$ vs $H$, when the condition $H_\omega \geq H_0$ of the nonlinearity was satisfied.

Figure 18 shows an example of a record of the nonlinear RFSE in **Bi** [69,73]. It can be seen that the left-hand edge of the nonlinear RFSE (red curve 2) is shifted into the range of weak fields because of the presence in the sample of a constant magnetic field created by the recified current. This shift increased with increasing $H_\omega$, in accordance with the theory suggested in Refs [70,73]. Furthermore, lines of nonlinear *RFSE* were observed in multiple magntic fields [70,73].

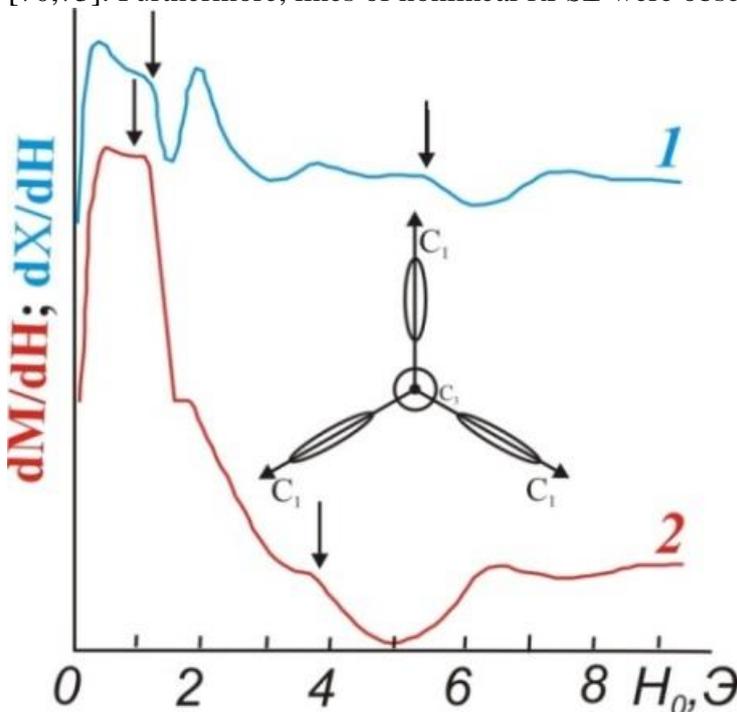

Figure 18. (Color online) Profiles of (1) linear and (2) nonlinear RFSE effect in **Bi** Amplitude of the HF field is $H_\omega = 0.1$ Oe. The arrows indicate the left-hand edges of the lines. The DC magnetic field is directed at the angle $13°$ toward the axis $C_2$; n//$C_3$ d=0.58mm [70,71].

At large amplitudes of the RF field, the rectified current may not disappear after the switching off the DC magnetic field, since it provides to be sufficient for the rectification. In other words, this means that in a metal plate upon excitation by an electromagnetic field with a large amplitude the "current" states, i.e. states in which a closed DC current flows in the sample can be implemented even in the absence of the external magnetic field [70-75].

An interpretation of the results of investigations into "current states" in different metals is presented in details in review [71]; therefore it is beyond the scope of this review, which is devoted to the fermi ology of metals.

## 3 De Haas-van Alphen effect

### 3.1 Quantum oscillations in $ZrB_{12}$

As has been noted in the Introduction, the other most popular effect for the study of FSs in metals are quantum oscillations of magnetic susceptibility (de Haas – van Alphen effect), and of the conductivity of metals (Shubnikov – de Haas effect). In these effects, the area of the extremal sections of the FSs is measured, which substantionally complicates the restoration of the FSs. Nevertheless, we consider below a number of results obtained recently in the case of high-temperature superconductors, without claiming the information available on this enormouse subject is complete [1,2,10]. As is known from the theory of the electronic structure of metals [74], part of the energy of electrons related with the motion in the plane perpendicular to the magnetic field is quantized in strong magnetic fields. In the quasiclassical case, the Lifshits-Onsager quantization condition takes on the following form [76]:

$$S(\varepsilon, p_z) = \frac{2\pi \hbar e H}{c} \times (n + \gamma) \quad n = 0,1,218) \ldots \quad (18).$$

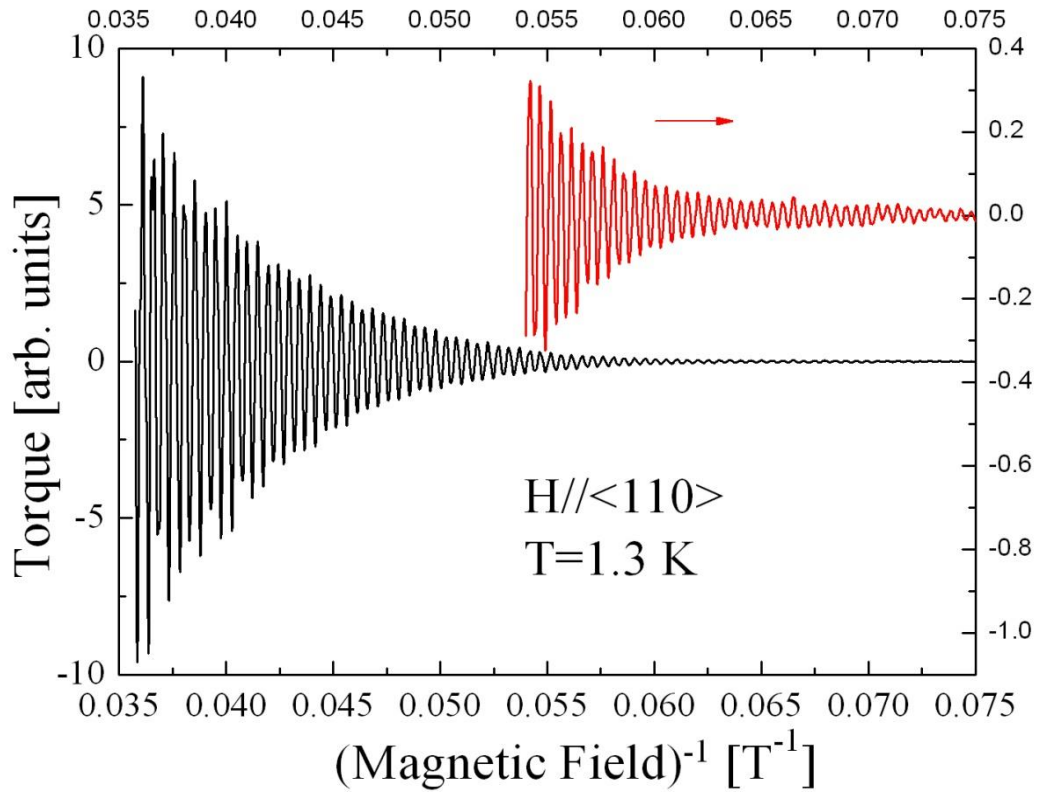

Figure 19, (Color online) Typical example of dHvA oscillations in $ZrB_{12}$ single crystals in a magnetic field oriented along the [100] axis [75].

Here, $H$ is the strength of the magnetic field along the z-axis; $S(\varepsilon, p_z)$ is the area of the section of the FS by the plane $p_z$ = const, perpendicular to H; $\gamma$=1/2, and e is the electron charge. The quantuzation of the electron energy leads to oscillations of the magnetic susceptebility $M$ and of the conductivity of metals in the reversed magnetic field 1/H, which makes it possible to determine the area of the extremal sections of the FSs $S_{extr}$:

$$S_{extr} = \frac{2\pi\hbar}{c\Delta H^{-1}}, \tag{19}$$

The frequency of the oscillations of the parallel magnetisation $M_\parallel$ is proportional to the surface area $S_F$ of the corresponding sheet of the FS: $F=(S_F\hbar c)/2\pi e$. In metals, the amplitde of oscillations (A(T,H) is described by the Lifsits-Kosevich formula [74]:

$$A \propto (H^{\frac{1}{2}}) \times \left|\frac{\partial^2 S_F}{\partial k^2}\right|^{-\frac{1}{2}} \times \frac{\frac{\alpha m_c T}{H}}{\sinh(\frac{\alpha m_c T}{H})} \times \exp(-(\frac{\alpha m_c T_D}{H})), \tag{20}$$

$$\alpha = \left(\frac{2\pi^2 c k_B m_0}{e\hbar}\right) = 14.693 T_D[K]$$

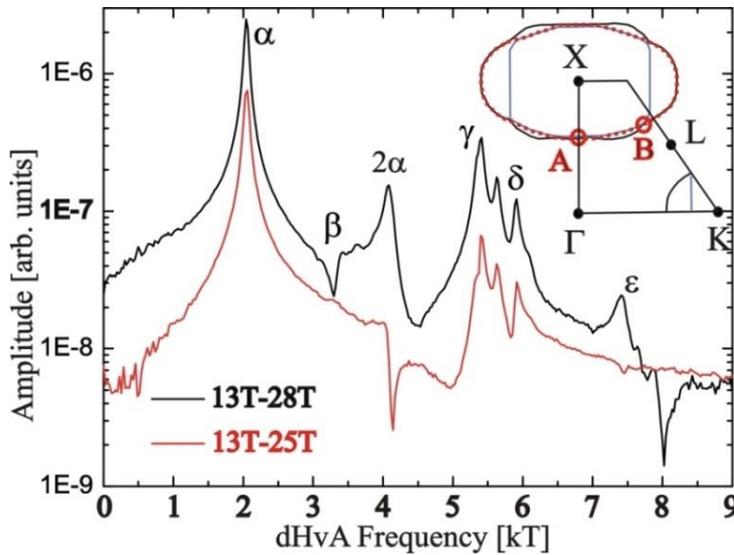

Figure 20. (Color online) Fourier expansion of quantum oscillations in $ZrB_{12}$ for a magnetic field directed along the [110] axis [75]. Inset shows, "the cubic box" type orbits.

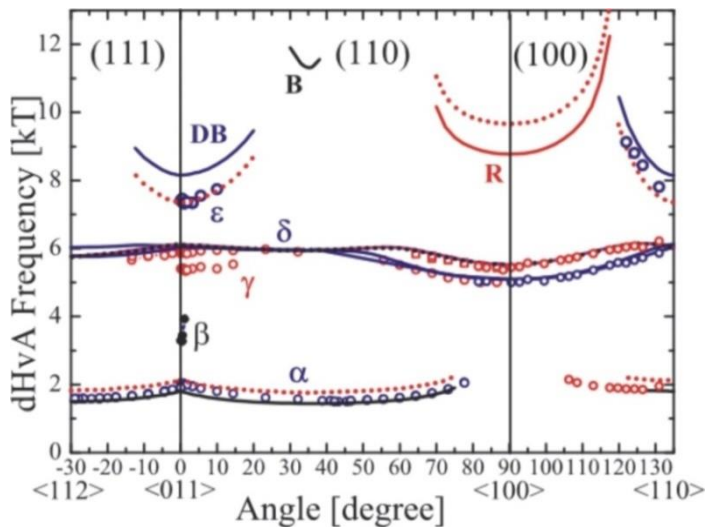

Figure 21. (Color online) Experimental results (points) for the anisotropy of the frequency of *dHvA* oscillations and theoretical curves, which correspond to extremal orbits on the FS of $ZrB_{12}$ "dogs bone" DB), "belly" (B), and "rosette" (R), respectively. The solid curves is the result of band strcture calculations of $ZrB_{12}$, for $E_F$ =0.16eV) [75].

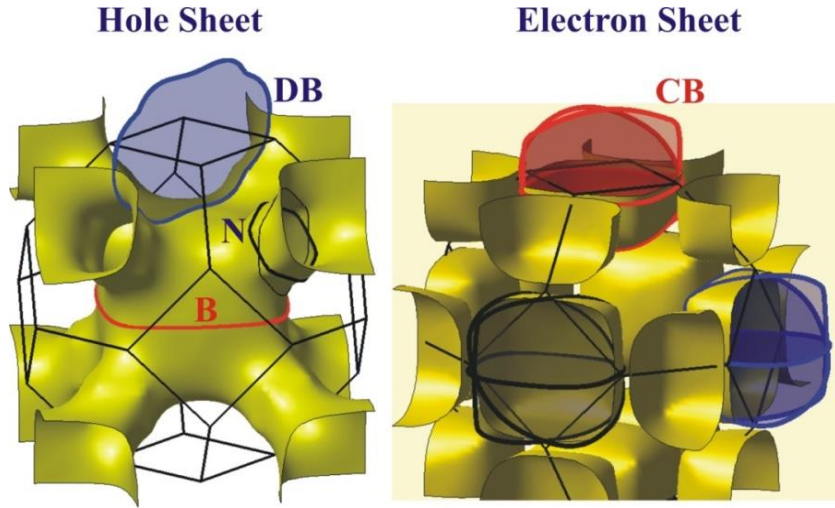

Figure 22. (Color online) Hole and electron sheets of the FS in the theoretical model of the FS of $ZrB_{12}$ [75]. Here, DB and CB notations correspond to "dogs bone" and "cubic box type orbits.

Here, $m^*_c = (\partial^2 E_F/\partial k^2)$ is the cyclotron effective mass of electrons, and $T_D$ is the Dingle temperature, which is inversely proportional to the scattering rate of electrons. Notice that $\beta$ and $\varepsilon$ peaks in Fourier expansion of these oscillations (Fig.20) appear only in strong magnetic fields because of magnetic breakdown of the "dogs bone" orbits and "cubic box" type orbits (see inset to Fig.22).

Equation (19) makes it possible to determine the frequency of oscillations, and correspondingly, the area (19) of the extremal cross-sections of the FS. As an example of studies of the *dHvA* effect, the quantum oscillations of $ZrB_{12}$ are presented in Fig.19 [75]. The solid curves in Fig.21 demonstrate a close agreement between the experiments *and ab initio* calculations of the energy-band electronic structure of ***$ZrB_{12}$***. The theoretical model of the FS of ***$ZrB_{12}$*** is shown in Fig.22 for the hole and electron sheets of the FS. The red curves in Fig.22 depict the orbits on the ***"neck"*** and on the "cubic box" black belts orbits. The minimum frequency on $\alpha$ – branch, between $1.2kT$ and $2\ kT$, corresponds to the orbits on the neck of the FS (orbit *N* in Fig.22). The branches near $6kT$ (see (Fig.20) are caused by the ***"cubic box"*** electron orbits. As a result of the conducted investigations, it was possible to identify all sections of the FS in ***$ZrB_{12}$***.

Some discrepancy between the experiment and calculations is caused by imperfection of the theoretical model, just as, strictly speaking, in the case of the FS of ***Ag*** [29] (see Section 2.3.). The above presented results demonstrate the advantages of the experimental techniques above the theoretical calculations of electronic structure of metals for restoration the FSs, which has been already discussed above (see Section 2.3).

The *dHvA* effect in the extreme case when $4\pi|(dM/dH-1|\leq 1$, leads to the formation of diamagnetic domains [76]. But this is a topic of a separate review [77].

### 3.2. Quantum oscillations in $YBa_2Cu_3O_{7-x}$

Another example of recently very successful studies of FSs in superconductors is quantum oscillations in single crystals of the high-temperature superconductor $YBa_2Cu_3O_{7-x}$ (YBCO) and $HgBa_2Cu_{4+x}$ (HBCO) [78-89]. In these papers, quantum oscillations of magnetic susceptibility, high-frequency, conductivity, magnetic moment, the Hall effect, and the Seebeck and Nernst effects have been studied in a wide range of magnetic fields (20-101T) and temperatures (1.1-30K.

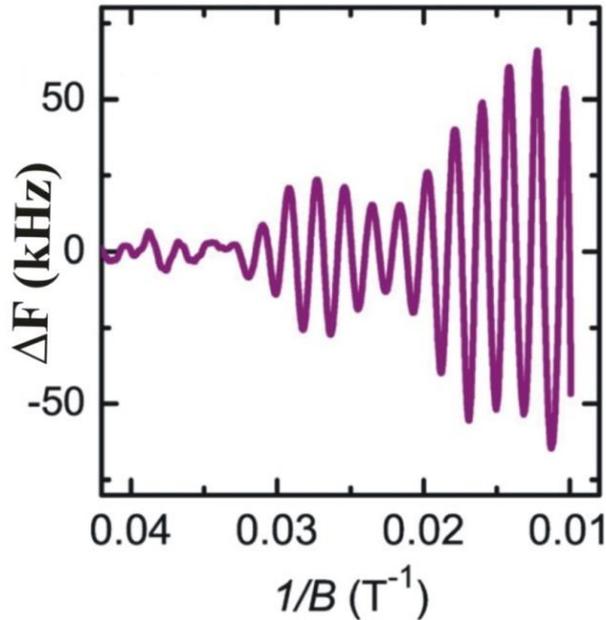

Figure 23. (Color online) Quantum oscillations in under doped $YBa_2Cu_3O_{7-x}$ with x =0.56 up to 101 T at 1.5 K versus reversed magnetic field 1/B [88].

Figure 23 presents an example of quantum oscillations of a single crystal YBCO (x=0.56), measured on the dependence of the frequency of a tunnel-diode generator on the reversed magnetic field [88]. The measurements were conducted using the same procedure as the measurements *HFSE*, based on the shift in resonance resonant frequency of auto dyne generator frequency depending on the magnetic field.

As a result of the investigations with TDO, quantum oscillations were observed up to 30 K, with a frequency oscillations which corresponds to small pockets on the FS of the YBCO, with a rather small effective mass $m^*=0.45\ m_0$ (see Fig. 23). A Fourier analysis of these oscillations (Fig.24) showed that they were caused by both electron and hole sheets of the FS. Indeed it was established in Refs [80-91], that the FS of the *YBCO* consists of one electron lozenge-like form of a pocket and two hole pockets in the form of deformed ellipsoids, in an accordance with the reconstruction of the FS caused by charge-density waves (CDWs) [88] (Fig.25.

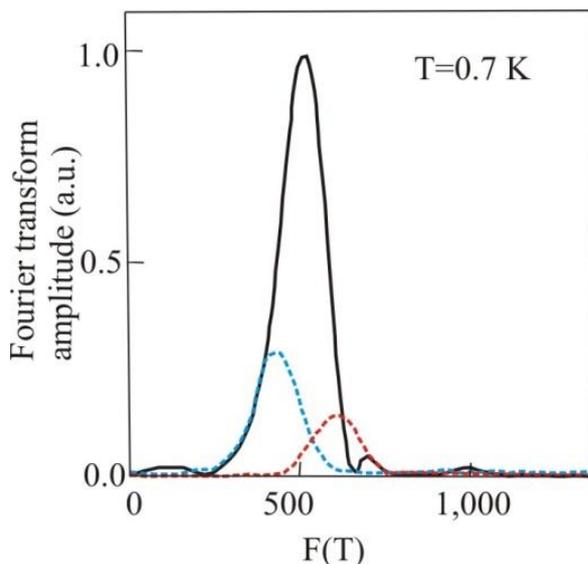

Figure 24 (Color online) Fourier expansions of quantum oscillations in the single crystal $YBa_2Cu_3O_{7-x}$

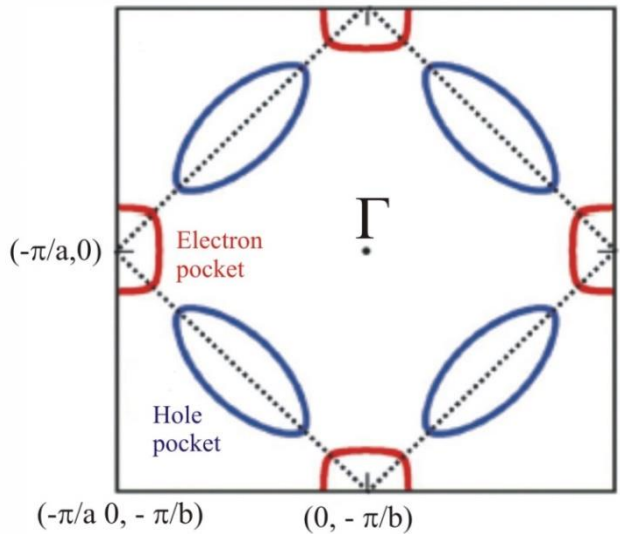

Figure 25. Color online), Sections of a reconstructed FS of YBCO, which consists of a lozenge-like electron pocket (red) and hole (blue) ellipsoid like sheets [88]. The dotted lines correspond to the anti-ferromagnetic Brillouin zone.

Apparently, it is possible that two of the ellipsoids in Fig.25 are related with two different sheets of the FS, which are caused by $CuO_2$ planes in the unit cell of the *YBCO* [88]. The third frequency of oscillations in Fig.24 is possibly due to the magnetic breakdown between the electron and hole sheets, of the FS, which is very likely in such strong magnetic fields, as 20-101T [88].

2. **Angle-resolved photo-electron spectroscopy**

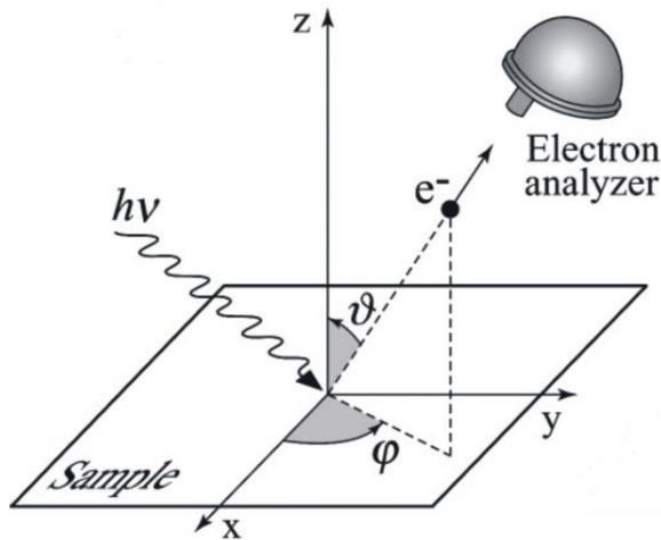

Figure 26. Color online), Diagram of an ARPES experiments in which the direction of the electron emission is determined by the angle *v* and the polar angle *φ*. The electron emission is registered by an electron energy analyzer [15].

This method of angle-resolved photoelectron spectroscopy (ARPES) occupies a special field among contemporary method of study on FSs of metals. A special future of this method of study the electronic structure of metals is the capacity to study low- lying energy level spectra of the metals located below the Fermi level [16,17].

The *ARPES* method is based on the phenomenon of the photo effect. Upon irradiation of the surface of a metal by ultraviolet radiation the emission of electrons of different energies occurs, which is registered in the form of a photoemission spectrum, and if using a sweeping over the angle, is recorded as an angle-resolved photoemission spectrum (Fig.26) [16-17]. In the *ARPES* experiments, the electrons emitted from the surface of the metal fall onto a hemispherical analyzer, whose lens directs the photoelectrons onto special multichannel plates. Upon further flight through the analyzer, the electron beam is upswept in energy in the plate perpendicular to the slit, and a two-dimensional spectrum forms on a 2D detector: the intensity of photoelectrons as a function of their energy and the angle of ejection. The sample and the detector are placed in an ultrahigh vacuum, since *ARPES* is a surface-sensitive technique. As the source of irradiation, either synchrotron emission (20-200 eV) or the emission of a helium plasma ~ 20 eV) is used, recently, modern lasers with quantum energy of 7 and 11 eV began to be used [15-17].

The *ARPES* method is based on a not-entirely substantiated assumption that an electron, when being emitted from a crystal into a vacuum, preserves its momentum and energy, i.e. it is assumed that on leaving the metal the quasi-momentum *ℏk* of electron passes into real momentum in a free space **p**= *ℏk+ℏG*, where **G** is the vector of the reciprocal lattice of the crystal. In this case, the energy of the electron is retained: $E_{kin}=\hbar v-\varphi-|E_b|$, where $E_b$ is the electron binding energy in the metal.

The intensity of the photocurrent is proportional to the special function multiplied by the Fermi distribution function: *I(k,ω) = A(kω) f(ω)*. Without taking into account the interaction, we have: *A(k,ω) δ[ω-ε(k)],* and the electronic spectrum is determined by the law of the energy-band distribution *ε(k)* in the metal. Allowance for the interaction strongly complicates the behavior of the special function:

$$A(k,\omega) = \frac{\Sigma''(\omega)}{\pi(\omega-\epsilon(k)-\Sigma'(\omega)^2+\Sigma''(\omega)^2} ,\qquad(21)$$

Here *Σ'ω+iΣ''is* the quasiparticle self-energy, which reflects the interaction in the metal. Generally speaking, there has been no such substantiation of the effect of *ARPES* to date; therefore, as a rule, different approximations are used [16-17]. It should be noted that *ARPES* is a surface-sensitive method, since the mean free path of photoelectrons greatly depends on energy, with a minimum at 2-5 Å. Therefore, it is not entirely obvious to which extent the *ARPES* spectrum reflects the real electronic structure of the bulk of the metal, rather than the structure of surface states. Thus, it was shown in Refs [16-17] that the photoemission in cuprates occurs from the depth on the order of two unit cell constants, i.e. ~ 15 Å.

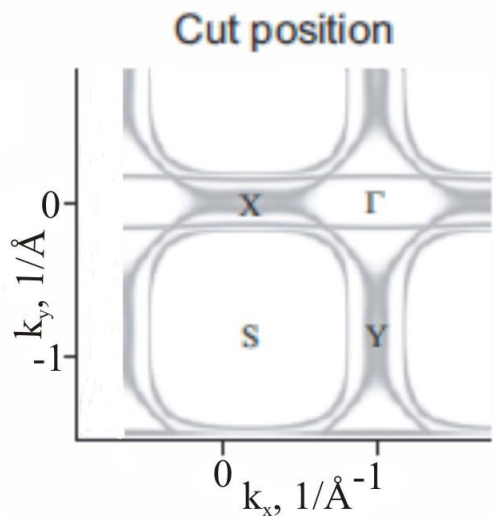

Figure 27. Color online), Electron-band structure of a de-twinned single crystal of $YBa_2Cu_3O_{7-x}$ [92].

Figure 27 displays a model of the section of the FS of a YBCO single crystal from *ARPES*, which consists of a pairs of outlines around point **S** in the Brillouin zone, corresponding to the connected and anti-connected sheets of the FS, restored with the aid of *ARPES* by the application of circularly polarized ultraviolet radiation [90].

In these experiments, it was possible to localize the position of surface and bulk states in the YBCO, where the $CuO_2$ layer nearest to the surface is considered to be localized, and the next bilayer is considered as lying in the bulk. At present, apparently, there is a fundamental discrepancy between *ARPES* data for FSs and the results of quantum oscillations [78-89] (see Fig. 27 and Fig. 25). It should also be noted that ARPES studies were carried out in zero fields [16,19,92]. Nevertheless, it seems that the existing nonconformity requires careful consideration.

Summarizing the results of investigations of quantum oscillations and *ARPES* in YBCO, note the following problems [88]: (1) a sharp decrease of the dimensions of the FS was discovered upon going over from the under doped to over doped regime and a rapid decrease in the Femi surface velocity in the case of low doping; (2) *ARPES* reveals the transformation (collapse) of the large FS on the over doped side into unconnected arcs of the FS; (3) quantum oscillations are explained by the magnetic breakdown between the hole and electron pockets of the FS.

Note in conclusions that investigations of the FS of $YBa_2Cu_3O_{7-x}$ are beyond the scope of this review, which is devoted to the Fermi ology of pure metals.

## 5. Conclusion

In this review, we tried to gather together the results of studies of the Fermi surface of pure metals performed with the aid of high-frequency size effects. We have shown, that the high accuracy of these methods using examples of studies that did not enter into other reviews.

We have shown that high-frequency size effects are very sensitive and efficient methods for studying FSs and mechanisms of electron scattering and excitation of electromagnetic waves in metals. Unfortunately, such studies are presently very rare because of nesessity of using high purity metals with large mean-free path of electrons. At the same time the, application of *HFSE* for studies of the FSs of nanostructured objects with a thickness of several micrometers can prove to be very efficient and useful.

It has been shown with the examples of the electronic structure of high-temperature superconductors, that the investigations of the FSs of metals are very important and fascinating problem in the study of the electronic structure of metals.

Acknowledgments. The author is grateful to V.T. Dolgopolov and V.G. Penschansky for valuable remarks concerning this review. The author is grateful to M.A. Harutunian, V.V. Bondarev, V.M. MacInnes, R. Stuby, P.-A Probst, *Collet*, B, K. Saermark, J.Lebech, C.K. Bak, and J. van der Maas, for their collaboration. The study was supported by the Russian Academy of Sciences program: Vital Problems of Low-Temperature Physics of the Russian Academy of Sciences. Part of these studies were carried out at the Institut de Physique Experimentale, University de Lausanne, Switzerland and was supported by the Swiss National Sience Foundtion for Scientific research, the author expresses his deep gratitude to these foundations and organizations.